# Surface forces and frictional properties of adsorbed bio-based cationic polysaccharide thin films in salted aqueous medium

*Lionel Bureau[1], Simon Trouillé[2] and Gustavo S. Luengo[2]*

[1]Université Grenoble Alpes, CNRS, LIPhy, 38000, Grenoble, France
[2]L'Oréal Research and Innovation, Aulnay-Sous-Bois, France

Inter-surface forces mediated by polymer films are important in a range of technological and industrial situations. In cosmetics, applications such as hair conditioning typically rely on the adsorption of polyelectrolyte films onto the charged surface of hair fibers, whose contact mechanics and tribological properties are central in determining the final sensorial perceptions associated with the cosmetic treatment. A major current challenge to be tackled by the cosmetic industry is to design high-performance products employing bio-sourced polyelectrolytes, with the aim of achieving eco-sustainable processes and products. In this context, the present study focuses on the mechanical properties of thin films obtained by adsorption from solution of fungal chitosan onto negatively charged mica surfaces. We use a Surface Forces Apparatus allowing for the simultaneous measurement of film thickness and friction force as a function of the applied normal load and shear velocity. We show that, in aqueous medium at an ionic strength of 40 mM, adsorbed films of chitosan give rise to repulsive inter-surface forces whose range, comparable to the Flory radius of the macromolecules, increases with the polymer molecular weight. In addition, the adsorbed layers are found to behave, under compressive forces, as pseudo-brushes of neutral polymers. Finally, we show that under shear forces, chitosan layers exhibit a transition from a low to a high friction regime under increasing confinement.

# Introduction

In cosmetics, the replacement of petrochemical-derived polymers in consumer formulations with natural, bio-based and/or biodegradable polymers has become a focus of research and development as material component sourcing and fate play an increasingly important role in product design [1]. These polymers frequently exhibit different topologies, raw material compositions, and solution behaviors from their petroleum-based counterparts. Comprehensive knowledge of the behavioral variations amongst polymers is necessary for an efficient and effective reformulation that preserves performances similar to that of the current goods.

Some of these polymers are used on rinsing applications (as shampoos or cleansers) to improve cosmetic sensorial attributes as slipperiness or touch, which therefore requires polymer affinity towards the cosmetic substrate (hair or skin). This attractive force towards the substrate allows for thin polymer monolayers to be adsorbed and support water flow, providing good lubrication in wet but also in dry conditions. They can provide skin protection properties, due to their functional surface properties and their emollient smooth feel.

In the case of hair, its surface is composed of scales arranged in roof-tile structure that form the cuticle layer (~5μm thickness) surrounding the inner part of the fiber called the cortex. On this outermost surface, chemically bound lipids coexist with areas of protein (keratin) exposed in a random manner at a density depending on the degree of weathering or oxidation of the hair [2]. The skin substrate is also complex and it is the stratum corneum, the outermost layer, that forms the external barrier. Here ceramides are also bound to flat keratinized cells (corneocytes) providing a complex geometrical morphology (*microrelief*).

At the common pH of classical formulations (5-7) keratin aminoacid groups and in particular sulfonates and acids are ionized and provide an overall negative charge to the surface. This implies that electrostatic forces are key driving forces for adsorption of charged ingredients such as cationic polymers. The reader is invited to refer to the dedicated literature that describes in more detail many of the cosmetic substrate characteristics [3,4].

The adsorption properties of the main formulation components, such as surfactants and polymers in hair care, have been extensively studied using surface-sensitive techniques such as ellipsometry or Quartz Crystal Microbalance, providing a glimpse of the characteristics of the thin film deposits in term of thickness and water content. Charged surfaces used for these studies are typically silica or mica (alone or modified at different levels of complexity), which are good models of a weathered hair surface [5,6]. In parallel, mean field calculations and molecular dynamic simulations have helped us understand the optimal polymer structural conformations and associations with surfactant that can better attach to the complex patterned structure of hair [7–9]. While these approaches have helped use predict the behavior of polymer chains during flow and to achieve a numerical estimation of their lubrication potential, rigorous experimental data are still scarce, in particular for bio-based polymers.

Characterizing and understanding intersurface and frictional forces mediated by adsorbed or grafted layers of solvated polymers has been an active research area since the late 80s [10–13]. A particular focus has been put on interactions between surfaces decorated by charged polymers [14,15], notably in the context of (bio)lubrication [16–21]. In this context, the properties of chitosan, the most abundant bio-sourced cationic polymer, has been surprisingly little studied: Claesson and Ninham [22] performed a pioneer study of interaction forces in the presence of surface-adsorbed chitosan at low ionic strength and various pH, Kampf et al. [23] further investigated the frictional properties of adsorbed chitosan layers, while more recently Kan et al. studied adhesive forces between chitosan films as a function of counterion valency [24,25], Xian et al. investigated the adhesive behavior of PEGylated chitosan layers [26], and Kwon et al. probed the hydrophobic interactions between chitosan layers and various self-assembled monolayers [27]. In the present study, we report Surface Forces Apparatus experiments investigating the effect of chitosan molecular weight in controlling normal and frictional forces between adsorbed layers under a high ionic strength typical of shampoo or hair conditioning conditions. Our study focuses on chitosan of fungal origin as a model for non-animal biosourced polyelectrolytes. This work complements previous studies performed with a single chain length [22,23]: it suggests that "salted" adsorbed chitosan behaves similarly to neutral polymers, and we demonstrate that theoretical models developed for brushes of neutral polymers can be used to rationalize our observations concerning the range and magnitude of surface and frictional forces.

# Materials and Methods

**Reagents.** All aqueous solutions were prepared in MilliQ water (18.2 MΩ.cm resistivity). Acetic acid (AcOH, 99.7%, Sigma-Aldrich, France) and sodium hydroxide (pellets, GPR rectapur, VWR, France) were used as received. Chitosan samples from fungal origin (Aspergillus Niger) were provided in dry form by L'Oréal. Molecular weights were determined by size exclusion chromatography and the degree of de-acetylation was obtained by $^1$H and $^{13}$C NMR characterizations by L'Oréal. Properties are reported in Table 1.

*Table 1. Properties of chitosan samples*

| Sample # | Mn (Da) | Mw (Da) | PDI | De-acetylation (%) |
|---|---|---|---|---|
| S1 | 15400 | 29700 | 1.93 | 97.3 |
| S2 | 15500 | 39400 | 2.54 | 95 |
| S3 | 39800 | 102600 | 2.58 | 92 |
| S4 | 72600 | 133300 | 1.84 | 91.7 |
| S5 | 116000 | 262600 | 2.26 | 95.7 |

It should be noted that the chitosan samples under study here display rough properties compared to synthetic polycationic systems. Size distributions are broad, and we expect *e.g.* sample to sample variations in purity, as chitosan samples from fungal origin are known to exhibit, in particular, residual β-glucan impurities at a fraction that may depend on the details of the chitosan extraction process [28,29]. In spite of such limitations, we show in what follows that rational trends can still be clearly identified regarding the role of chitosan molecular weight.

**Solution conditions.** All chitosan samples were diluted at a mass concentration of 2 g.L$^{-1}$ in MilliQ water containing 4 mM of acetic acid, 1.5 mM of NaOH (pH=4.5), and 40 mM of NaCl. We observed a small insoluble fraction in our solutions, most probably associated with β-glucan residues as mentioned above. Such fraction was left to sediment overnight, forming a turbid layer at the bottom of the vials (occupying about 2-3% of the solution volume), and only the clear supernatant was used for experiments.

**Substrate preparation.** Muscovite mica (grade V-1) was obtained from Neyco (France), cleaved and scissor-cut into sheets of approximately 1 cm² and thickness 10-20 µm. One side of the sheets was evaporated with a 50nm-thick layer of silver (99.99%, Neyco, France) on a magnetron-sputtering machine. Mica sheets were then glued, silvered side down, with NOA 81 UV-setting adhesive, onto cylindrical optical lenses (Thorlabs, France), thus imposing a macroscopic curvature to the mica, of radius R=1 cm. Once glued, mica sheets were cleaved with scotch tape in a laminar flow cabinet and installed in the instrument immediately prior to measurements.

**Surface Forces Apparatus measurements.** Surface and frictional forces were measured on a home-built Surface Forces Apparatus (SFA). The instrument is sketched on Fig. 1; it is an updated version of a previously developed instrument [30]. In brief, the bottom mica surface is installed in a small liquid trough of volume 3 mL. The trough is mounted on a 2-axis piezo-actuator (Physik Instrumente P-733.3 positioner with E503 amplifier, E509 signal conditioner and E517 digital interface) allowing for displacements of the substrate in the normal (z) direction with Angstrom accuracy over a 10µm stroke and in the lateral (x) direction with nm accuracy over 100 µm. The top mica substrate is installed on a 2-axis force sensor made of 2 crossed double-cantilever springs whose deflections are measured by means of contactless capacitive displacement sensors (50µm-range sensors CS005 with DT6220/DL6230 electronics, Micro-Epsilon, France). The static resolution of the displacement sensors of 2.5 Å combines with the 600 N.m$^{-1}$ stiffness of each dual-cantilever spring to yield a nominal force resolution in both x and z direction of 0.15 µN. In practice, the effective force resolution is limited to about 0.5 µN by residual mechanical vibrations. The intersurface distance between the mica sheets is measured optically, with Å resolution, by analyzing the spectral position of the fringes of equal chromatic order (FECO, observed with an Acton Spectrapro 2500i spectrometer equipped with a Pixis 400B camera) resulting from white light illumination (Schott KL2500 LED) of the Fabry-Perot cavity formed by the two back-silvered mica sheets. Real-time computation of the intersurface distance is performed using the multilayer matrix method described in previous studies [30,31].
In a typical experiment, the two freshly cleaved mica sheets were first brought into adhesive contact in air to measure the total thickness of the mica substrates, then separated in air to determine their respective thicknesses, as described in [30]. Next, 3mL of chitosan solution, filtered through a 0.2 µm membrane to remove possible dust particles, were introduced into the liquid trough, and both mica surfaces, separated by about 2-3 mm, were left immersed in the solution for 15 minutes for polymer adsorption to proceed. Following adsorption, the chitosan solution was removed, without drying the substrates, and replaced by buffer (i.e. MilliQ water with 4 mM acetic acid, 1.5 mM of NaOH and 40 mM of NaCl). Mica substrates were left for 15 min in buffer to remove non-adsorbed material, and this rinsing step was repeated twice, using fresh buffer each time. After rinsing, buffer was replaced once more, and the surfaces left to equilibrate for 30 min before starting measurements at room temperature (22±1°C).

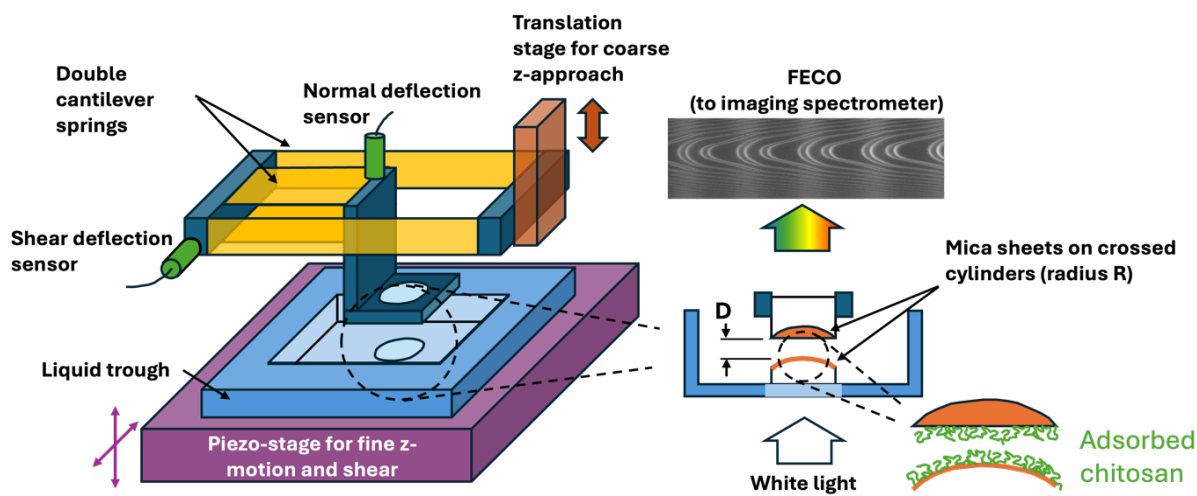

*Fig. 1. Sketch of the surface forces apparatus.*

Interaction forces were first characterized by monitoring the normal force (F) as a function of intersurface distance (D) during approach and separation of the substrates. Such an approach/retract cycle was performed using a velocity of 1 nm.s$^{-1}$ of the z-axis piezo stage, this velocity ensuring that the magnitude of the hydrodynamic Reynolds drag is kept below the force resolution of the instrument and is thus negligible in the overall measured forces. After separation, surfaces were left to relax for 15 min before being approached again and brought under a stepwise increasing normal force F, up to F=1.2 mN. At each normal force step, the bottom surface was translated laterally, forth and back, using the x-axis piezo stage, at a velocity of V=1 µm.s$^{-1}$, over a distance of 10 µm. The normal force signal was used as the input of a proportional-integral-differential feedback loop of the z-displacement axis, to keep the normal force constant during shear motion, while monitoring the resulting tangential (friction) force T at steady-state, along with the distance D. When reaching repulsive normal forces on the order of 400 µN, the friction force was measured as a function of shear velocity in the range 0.03-30 µm.s$^{-1}$.

# Results

**Normal forces**. A typical approach/separation force curve measured after chitosan adsorption (sample S4) is shown on Fig. 2a. Upon approaching the surfaces, we detect repulsive forces, setting in at intersurface distances of several tens of nm and increasing monotonously as the distance D decreases further. When separating the surfaces, the F(D) curve is found to lie below that measured upon approach, indicating hysteresis in the approach/separation cycle. Surface separation takes place without any evidence for adhesive forces. Such qualitative features are common to all the force curves measured, irrespective of the molecular weight of the adsorbed chitosan (Fig. S1 of Supplementary Information).

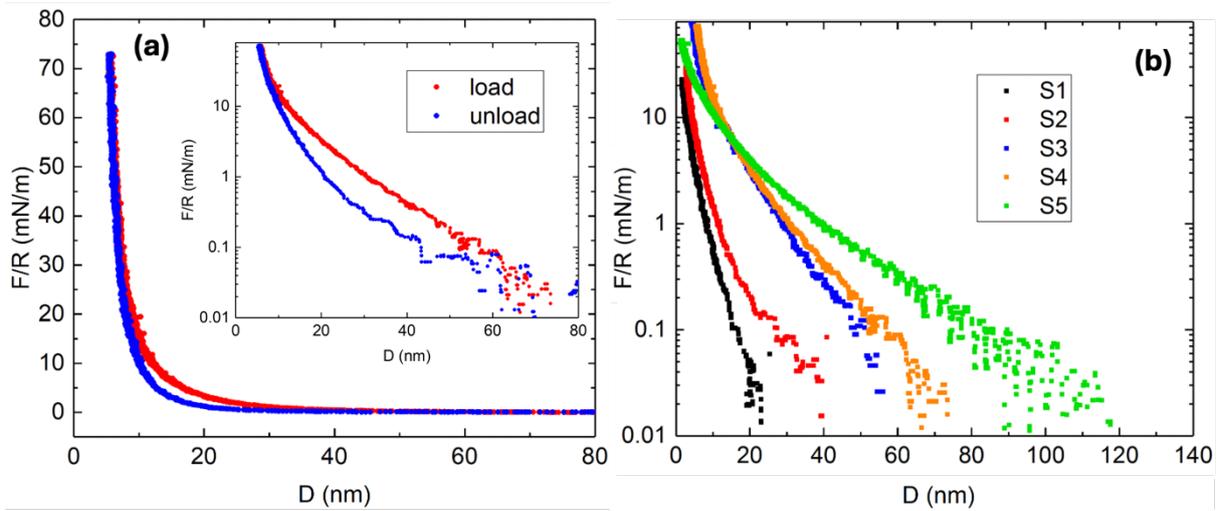

*Fig. 2. (a) Normalized normal force (F/R, with R the radius) as a function of distance (D) measured after adsorption of sample S4, upon approach (red dots) and separation (blue dots). Insert: same dataset with F/R represented on a log scale. (b) F/R vs D during approach only, for all chitosan samples (S1: black, S2: red, S3: blue, S4: orange, S5: green).*

The impact of chitosan properties is illustrated on Fig. 2b, which compares force curves measured upon approach for the 5 different samples studied. The distance at which repulsive interactions build up is seen to gradually increase from ~20 nm to ~100 nm for sample S1 to S5.

**Friction.** Fig. 3a illustrates a typical time trace of the friction force measured on sample S2, along with the imposed shear motion. Fig. 3b shows the evolution of the steady-state value of the friction force, measured at V=1 $\mu$m.s$^{-1}$, as a function of normal load, for all chitosan samples. We qualitatively observe, for all samples, that the friction force is of low magnitude at low loads and increases non-linearly up to typically F=200-400 $\mu$N, beyond which it increases linearly with F (Fig. 3b). In addition, we note that in the low friction regime, the friction force relaxes down to zero upon cessation of shear, whereas at higher normal loads the shear force relaxes down to a finite value (Fig. 3a).

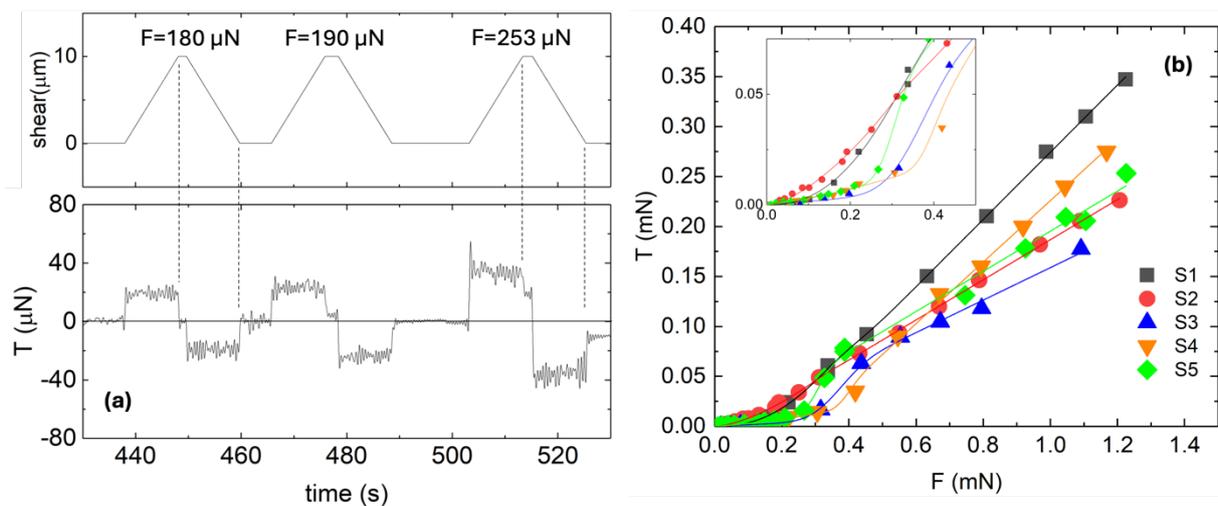

*Fig. 3. (a) Top: applied shear motion as a function of time, for 3 different normal loads as indicated above the time trace. Bottom: shear (friction) force as a function of time, measured for sample S2. Vertical dashed lines mark 4 "shear stop" events and the corresponding shear*

*force relaxation to zero (left) and to non-zero level (right). (b) steady-state friction force (T) as a function of normal load (F), measured at a shear velocity of V=1 μm.s⁻¹. Symbols are experimental data, lines are guides for the eye. Inset: zoom on data at F<0.4 mN.*

To better evidence the different frictional regimes exhibited by the confined chitosan layers, we compute a local friction coefficient[i] from each dataset shown on Fig. 3, defined as μ=dT/dF. We plot the evolution of the friction coefficient with normal load F on Fig. 4a. It can be seen that, except for sample S2, the friction coefficient of the sheared confined films is typically low, in the range μ=0.02-0.05, at low normal loads (F<100-300 μN), then increases markedly before stabilizing to a high (μ=0.15-0.35) plateau value corresponding to the linear regime visible on Fig. 3 at higher loads. Sample S2, though exhibiting higher values of μ at low loads, displays the same qualitative transition from low to high friction as F is increased. Furthermore, plotting μ as a function of intersurface distance D (Fig. 4b) reveals that such a transition from low to high friction happens rather sharply when adsorbed films are confined to thicknesses below 4-8 nm depending on samples. A comparison of the friction coefficients, either in the low or high friction regime, between the various chitosan samples does not reveal any systematic trend, suggesting that the friction coefficient is not dependent on chitosan molecular weight.

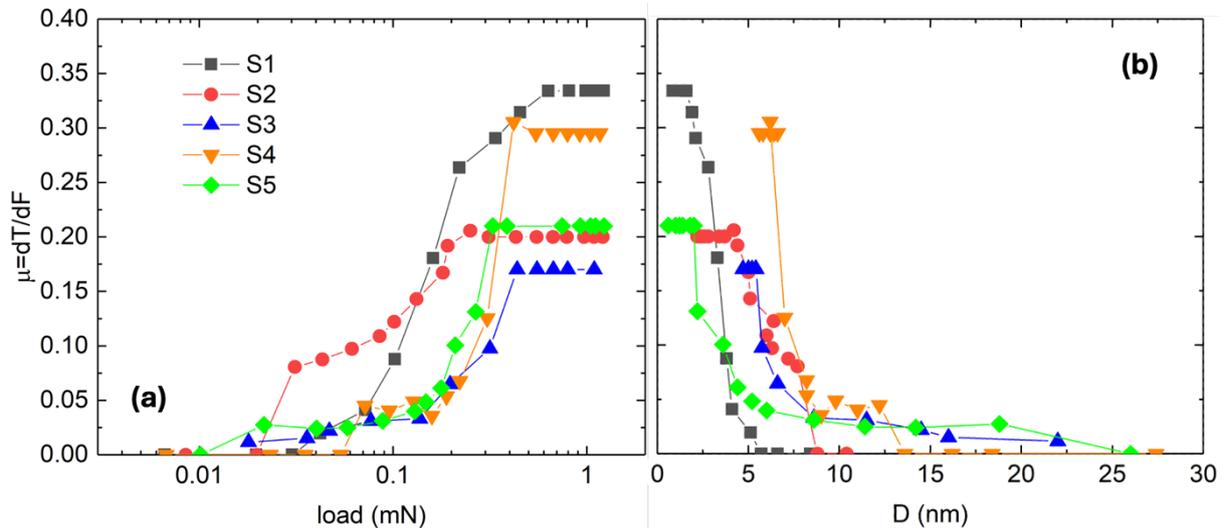

*Fig. 4. (a) Friction coefficient at V=1μm.s-1 as a function of (a) normal load, and (b) intersurface distance, for the five chitosan samples. Data points displayed with μ=0 correspond to friction forces that were below the experimental resolution of the SFA.*

The friction coefficient, μ=T/F, measured at F~400 μN, is observed to depend on shear velocity V, as shown on Fig. 5: qualitatively, all chitosan samples exhibit a μ(V) response displaying a shallow minimum at low velocity, in the range 0.1-1 μm.s⁻¹, beyond which a quasi-logarithmic increase of μ with V is observed. The slope of this logarithmic regime appears to depend on chitosan molecular weight, with longer chains displaying a stronger log-dependence on velocity.

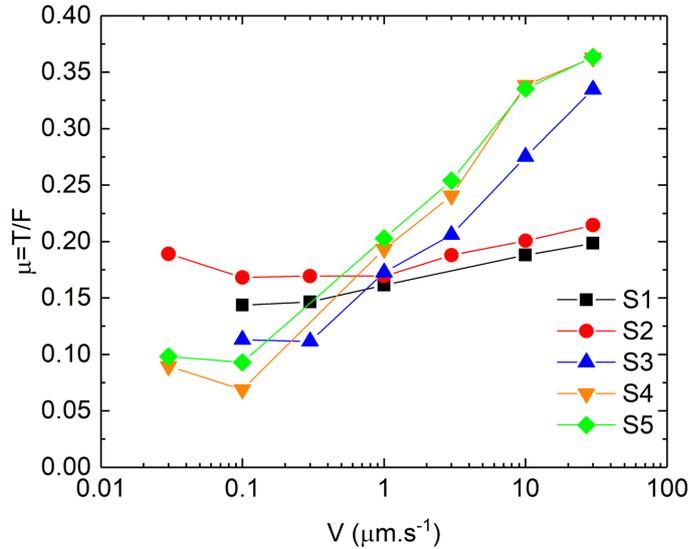

*Fig. 5. Friction coefficient as a function of shear velocity V.*

# Discussion

**Normal forces.** The purely repulsive interactions measured upon approaching and separating the surfaces are in qualitative agreement with previous SFA studies [22,23] performed in the presence of adsorbed chitosan films and at pH well below the pKa of chitosan (~6.5). It is however in contrast with a recent work reporting strong and contact-time dependent adhesion upon separation of mica substrates onto which chitosan was adsorbed [26]. We note that the latter study was conducted by adsorbing chitosan from very dilute solutions (0.001 w%) compared to other studies (0.01% [22], 0.065% [23], and 0.2% in the present work). Such a low mass concentration might yield to much less dense and thinner adsorbed layers in the work by Xiang et al. [26], possibly not screening mica/mica short-ranged attractive interactions responsible for adhesion.

In their pioneer study, performed in aqueous media of low ionic strength, Claesson and Ninham [22] attribute repulsive interactions to a combination of electrostatic double-layer forces at large surface separations and steric repulsion of the compressed chitosan layers at intersurface distances below about 10 nm. Kampf et al. [23], who studied a chitosan sample of molecular weight comparable to that of Claesson and Ninham (60-70 kDa with high deacetylation degree), associate the measured repulsive forces to solely steric effects setting in below distances on the order of 20 nm, electrostatic interactions being essentially negligible in their experiment. Such a steric origin of interactions between chitosan-coated surfaces was discussed on a qualitative basis by both groups of authors. In what follows, we attempt to go further in the interpretation of repulsive forces by analyzing the measured force curves and their dependence on chitosan molecular weight.

In their scaling theory of polyelectrolyte adsorption onto oppositely charged substrates, Dobrynin et al. [32] predict how the structure of an adsorbed layer depends on the charge density, $\sigma$, of the substrate and on the Debye length $\lambda_D$. They show that an adsorbed polyelectrolyte layer undergoes a transition from quasi 2D, with macromolecules lying flat on the surface, to 3D when the surface charge density $\sigma$ increases up to a value $\sigma_e \approx f/A$, with $f$ the fraction of charged monomers and $A$ the area per monomer. Taking an average value of $f \approx 0.25$ (equal to the average de-acetylation degree of our samples, 0.95, times the fraction of effectively charged units due to Manning's condensation, which is evaluated to be about 30% at pH 4.5 from the report of Lupa et al. [33]), and taking the monomer area [33] $A \approx 0.4$ nm$^2$, we estimate $\sigma_e \approx 0.6$ nm$^{-2}$, which is to be compared with the surface charge density of mica resulting from its crystallography, $\sigma \approx 0.5$ nm$^{-2}$. Such comparable orders of magnitude strongly suggest that the adsorbed layers in our study are likely to be 3-dimentional. Moreover, Dobrynin et al. predict that, at high enough salt concentration, when the Debye length is comparable to the size of the electrostatic blob, the thickness of a 3D adsorbed layer should scale as the size of the chains in solution. At the ionic strength used here (~42 mM), we compute $\lambda_D \approx$ 1.5 nm, and we estimate the electrostatic blob size $D_e \approx a^{4/3} l_B^{-1/3} f^{-2/3} \approx$ 1.1 nm (with $a$=0.5 nm the monomer length [33], and the Bjerrum length $l_B = 0.7$ nm in water at room temperature), which suggests that our experimental conditions correspond to the above-described theoretical regime of layer thickness. Based on such estimates, we thus expect from the theory of Dobrynin et al. that steric repulsion between two adsorbed salted chitosan layers would set in at intersurface distances on the order of twice the (molecular weight-dependent) size of individual chains. We estimate[ii] the average Flory radius of the various chitosan samples as [34] $R_F \approx n^{1/5} a N^{3/5}$, with $n = 10$ the number of monomers in the persistent segment [33], and $N = M_w/M_0$ the number of monomers per chain with $M_0 = 180$ g.mol$^{-1}$ the monomer molecular weight [33]. Doing so, we obtain $R_F \approx$ 17, 20, 36, 42 and 63 nm for samples S1 to S5. As discussed below, these values are in good agreement with the distances at which normal forces become measurable for each type of sample (see Fig. 2b).

The above analysis suggests that chitosan layers adsorbed on mica and immersed in low pH aqueous medium at high salt concentration can be considered as adsorbed layers of neutral polymers, due to the strong screening of inter- and intrachain electrostatic interactions [32] at the ionic strength of ~40 mM used in our study. We therefore attempt, in the spirit of a previous study on salted polyelectrolyte brushes [35], to apply theories established for adsorbed neutral polymer layers in order to model the force/distance curves shown on Fig. 2b. The situation where two layers of neutral polymers are compressed by approaching the surfaces onto which they are adsorbed have been described theoretically by de Gennes [36] and Aubouy et al. [37] Both works reach the same result: when two such "pseudo-brushes", adsorbed onto planar substrates, are approached at intersurface distances z below twice their unperturbed thickness ($H_0$), the osmotic pressure associated with the compression of the layers reads, at low compression $z \leq 2H_0$:

$$\Pi(z) \approx \frac{\alpha kT}{z^3} \quad (1)$$

with α a numerical prefactor of order unity and *kT* the thermal energy. The repulsive force resulting from layer compression down to distance D in the crossed-cylinder geometry of our SFA experiments can then be computed, within the Derjaguin approximation, as:

$$\frac{F(D)}{2\pi R} \approx \int_D^{2H_0} \frac{\alpha kT}{z^3} dz \quad (2)$$

Which yields for the normalized force F/R:

$$\frac{F(D)}{R} \approx \alpha \pi \frac{kT}{(2H_0)^2} \left[ \left(\frac{2H_0}{D}\right)^2 - 1 \right] \quad (3)$$

We have used Eq. 3 to fit the force curves shown on Fig. 2b, leaving $H_0$ and α as free parameters. Since Eq. 1 corresponds to a low compression prediction, fits were restricted to data corresponding to F/R below 1 mN/m. When rescaling the experimental data according to Eq. 3, i.e. plotting $F \times (2H_0)^2/(\alpha \pi RkT)$ as a function of $D/(2H_0)$, using for each dataset the best fit values of $\alpha$ and $H_0$, we obtain a satisfactory collapse of all datasets on a master curve, as shown on Fig. 6a. Moreover, we observe that the functional form of Eq. 3 quantitatively describes the experimental force curves for the various chitosan samples. Inspecting the best fit values for parameters $H_0$ and α leading to the collapsed data of Fig. 6a, we note two salient features. The layer thickness, $H_0$, is proportional to $R_F$, the average Flory radius of the adsorbed chains (Fig. 6b). This is in excellent agreement with a layer thickness on the order of the size of the unperturbed chains in solution, as anticipated by Dobrynin et al. [32] The best fit values for α are, however, more difficult to rationalize: we observe that α strongly increases with the average molecular weight of chitosan (Fig. 6c), whereas this parameter is expected to be a numerical prefactor independent of chain length [36]. While the good agreement between $H_0$ and $R_F$ points to the fact that adsorbed chitosan layers are of the theoretically expected thickness, the dependence of α on $M_w$ indicates that the magnitude of the repulsive forces is not correctly accounted for by Eqs (1)-(3).

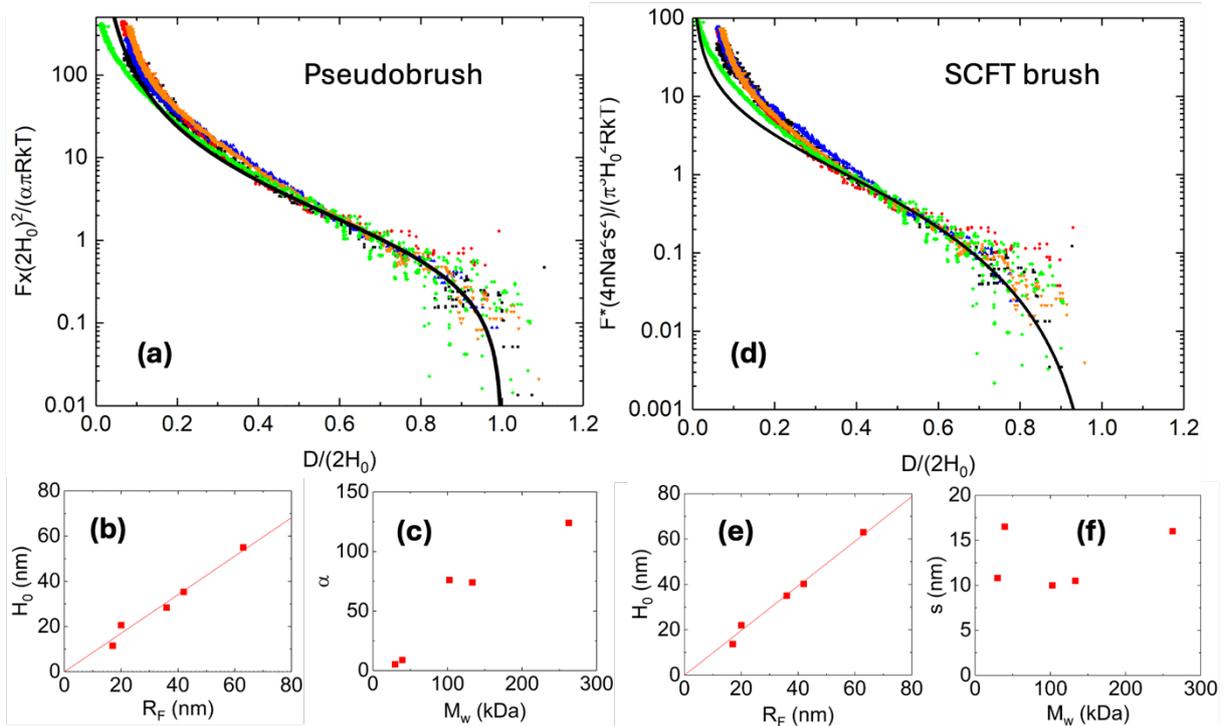

*Fig. 6. (a) Collapsed experimental force curves, with F and D adimentionalized according to Eq. 3 (symbols), compared to the theoretical pseudobrush prediction (line). (b) Best fit value for layer thickness $H_0$ (Eq. 3) as a function of the Flory radius $R_F$ of the chains (symbols). The line is the best linear fit of $H_0(R_F)$, with slope 0.85. (c) Dependence on chain molecular weight $M_w$ of the best fit value of parameter α. (d) Collapsed experimental force curves, with F and D adimentionalized according to Eq. 4 (symbols), compared to the theoretical prediction of the self-consistent mean field theory of polymer brushes (line). (e) Best fit value for brush thickness $H_0$ (Eq. 4) as a function of the Flory radius $R_F$ of the chains (symbols). The line is the best linear fit of $H_0(R_F)$, with slope 0.98. (f) Dependence on chain molecular weight $M_w$ of the best fit value of parameter s (Eq. 4).*

Following earlier studies of surface forces in the presence of adsorbed polyelectrolyte layers [15,38], we now attempt to describe the measured force curves by assuming that compressed adsorbed layers can be treated as polymer brushes, with protruding polymer tails and loops behaving as end-grafted chains. For this, we use the theoretical force-distance relationship established for salted hyaluronan brushes by Attili et al. [35] and adapted by the authors from self-consistent field theory (SCFT) of polymer brushes :

$$\frac{F(D)}{R} \approx \frac{\pi^3}{4} \frac{(2H_0)^2 kT}{nNa^2s^2} \left[ \frac{2H_0}{D} + \left(\frac{D}{2H_0}\right)^2 - \frac{1}{5}\left(\frac{D}{2H_0}\right)^5 - \frac{9}{5} \right] \qquad (4)$$

with $n, N$ and $a$ defined above, $H_0$ the layer thickness and $s$ the average distance between grafted chains. For polymer brushes, $H_0$ and $s$ are not independent [35]. For the case of adsorbed layers however, we cannot relate these two quantities, and we leave both $H_0$ and $s$ as free fitting parameters, as done in previous studies [15,38]. As above, we fit each individual force curve with Eq. 4, for F/R<1 mN/m, to obtain values of $H_0$ and $s$, and plot the rescaled force $F \times (4nNa^2s^2)/(\pi^3 R(2H_0)^2 kT)$ as a function of rescaled distance $D/(2H_0)$. As seen on Fig. 6d, such a rescaling again provides a satisfactory collapse of all the datasets. Furthermore, comparison of the experimental data with the functional form of Eq. 4 shows that force curves are properly described by polymer brush theory in the range $D/(2H_0) \geq 0.4$ (Fig. 6d). As for the pseudo-brush analysis above, $H_0$ is found to be proportional to $R_F$ (Fig. 6e) for the 5 molecular weights investigated. In addition, we find that, within scatter, $s$ is insensitive to chitosan molecular weight and lies in the range $s$=10-16 nm (Fig. 6f). Such an analysis strengthens the fact that adsorbed chitosan layers at high ionic strength cause steric intersurface repulsion setting in at distances on the order of the chain size $R_F$. This is in apparent contrast with the work of Tiraferri et al. [39], who report, at high salt and acidic pH, that chitosan of 170 kDa forms a layer of only ~1 nm in thickness. Such a quantitative discrepancy might stem from their use of chitosan with lower de-acytalation degree (70%) than ours, probed using techniques (quartz crystal microbalance and optical reflectometry) that are rather sensitive to the near-substrate and denser part of the layers, and not to the outermost low-density region of the layer, as is the case for SFA. We note in addition that, over the range of $M_w$ covered here, our observations are consistent with recent results from neutron reflectometry on adsorbed chitosan [40] showing that, while oligomeric chitosan of 3 kDa tends to adsorb "upright" onto the substrate, longer chains of 27 kDa form layers extending over a thickness of about 7-12 nm. Our results show that the repulsive forces arising from such layers can be, at least at low compression, described as the mechanical response of a neutral brush having an effective area per chain $s^2$ independent of chain length and in the range 100-256 nm$^2$ in the present study. Note that we reach a similar conclusion by using the Alexander-de Gennes brush model to fit our data, as shown in SI (Fig. S2). At strong layer compressions, $D/(2H_0) < 0.4$, we see that brush theory underestimates the measured forces. This might be attributed to the fact that the near-surface monomer volume fraction in the adsorbed layers is different from the parabolic profile of monodisperse brushes [32].

Finally, the hysteresis observed during compression/decompression of the layers is qualitatively consistent with previous SFA measurements of adsorbed polymer layers in good solvent [13]. It has been proposed by Raviv et al. [41] that such hysteresis stems from an increase in monomer/surface contact points upon layer compression, resulting in slowly-relaxing changes in the layer density profile, associated with a modification of the range of steric forces upon surface separation. Although we did not investigate the relaxation dynamics of the chitosan layers after their first compression, it is likely that such a mechanism is at play upon layer decompression in our measurements.

**Friction.** The transition from low to high friction when normal load is increased (Fig. 3b and 4) is consistent with several previous studies on polymer brushes [12] and adsorbed polymer layers in good solvents. Our observation is thus qualitatively in line with the frictional behavior of well-solvated (bio)polymer layers. Moreover, the shear force relaxations illustrated on Fig. 3a strongly suggests that such a low to high friction transition corresponds to a liquid-like (force relaxation to zero) to a solid-like (relaxation to finite force) frictional transition as the chitosan layers are compressed. The weak, quasi-logarithmic dependence of the friction coefficient on shear velocity (Fig. 5), which is observed in the high load regime, is typical of dry frictional contact involving solid polymers [42], and further supports a solid-like behavior of the strongly compressed chitosan layers.

Based on the above analysis of normal forces, which allows us to estimate an unperturbed thickness $H_0$ of the layers, we may further estimate the contact area formed between the compressed layers: indeed, the adsorbed layers being the most compliant part of the glue/mica/chitosan/mica/glue stack of materials in our experiments, we can evaluate[iii] the contact area as $S \approx 2\pi R\delta$ with $\delta = 2H_0 - D$ the load-dependent layer indentation. Doing so, we can compute the dependence of the frictional shear stress, $\tau = T/S$, as a function of the contact pressure $p = F/S$. We thus evidence that $\tau$ displays a non-trivial dependence on $p$, with a (low friction) regime where the shear stress increases non-linearly with pressure, followed by a (high friction) regime with a quasi-linear increase of $\tau$ with $p$ (Fig. 7a). This shows that the frictional properties of the compressed chitosan layers are not merely controlled by the contact area, and that the magnitude of the frictional stress depends on the level of confinement of the adsorbed layers. This is better demonstrated when plotting the shear stress as a function of the compression ratio $2H_0/D$, as shown on Fig. 7b: the frictional stress is seen to rise by about 2 decades (from below 1 to ~100 kPa) as the compression ratio increases, with a "liquid-to-solid-like" transition region in the range $2H_0/D \approx 60$ for sample S5, ~12 for samples S3 and S4, and ~7 for S1 (sample S2 displays unmeasurable shear stresses for compression ratios below 5, followed by a gradual increase in $\tau$ with compression). This overall trend indicates that chitosan layers of higher molecular weights can sustain larger compressive strains before exhibiting solid-like friction. Moreover, it appears that at a given compression ratio, in the low friction regime, the shear stress tends to be lower for chitosan chains of larger molecular weights.

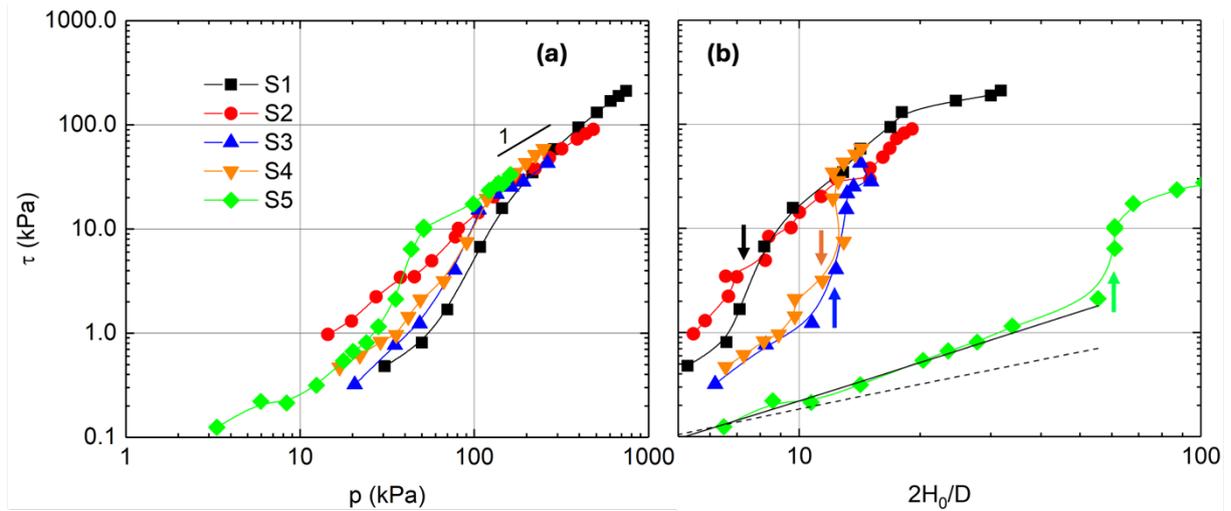

Fig. 7. (a) Frictional shear stress $\tau$ as a function of contact pressure $p$. The short black line indicates a power law with exponent 1 (i.e. linear increase). Solid colored lines are guide for the eye. (b) $\tau$ as a function of compression ratio $2H_0/D$. Arrows of the symbol colors mark the approximate position of the low-to-high friction transition for the various samples. The dashed line represents the prediction for the shear stress based on the model by Klein [45], with $\eta \approx 0.01\ Pa.s$ and $s = 7\ nm$. The full line is the theoretical prediction using Eq. 11, with $\eta \approx 0.011\ Pa.s$ and $s = 7\ nm$.

The brush-like behavior described in the above discussion of normal intersurface forces prompted us to use similar concepts to account for the friction of confined chitosan layers. In his review of forces between polymer-decorated surfaces, Klein [45] proposed a model describing the frictional stress between brush-bearing surfaces. The model, based on scaling arguments for the frictional drag due to interpenetrating brushes in good solvent, predicts for the shear stress between two compressed *planar* brushes:

$$\sigma_b \approx \frac{6\pi\eta_s V}{s}\left(\frac{2H_0}{h}\right)^{\frac{7}{4}} \tag{5}$$

with $\eta_s$ the solvent viscosity, $V$ the shear velocity, $h$ the distance between the grafted surfaces, and $s$ the distance between end-tethered chains. Eq. 5 holds for planar, uniformly compressed brushes. It thus cannot be directly compared to the friction stress $\tau$ determined above from the experimental measurements, since in the sphere/flat geometry of the SFA, the compression of the polymer layers decreases from its maximum value at the point of closest approach $D$ towards zero at the edge of the contact area. To compare Klein's prediction to our data, we need to compute the theoretical shear force T by integrating eq. (5) over the contact area, and then divide it by the contact area to compare with $\tau$. Within the Derjaguin approximation, we can write the theoretical shear force as:

$$T \approx 2\pi R \int_D^{2H_0} \sigma_b(h)\,dh \approx 2\pi R \frac{6\pi\eta_s V}{s}\int_D^{2H_0}\left(\frac{2H_0}{h}\right)^{\frac{7}{4}}dh \tag{6}$$

We thus obtain for the friction force:

$$T \approx \frac{32\pi^2 R H_0 \eta_s V}{s}\left[\left(\frac{2H_0}{D}\right)^{\frac{3}{4}} - 1\right] \tag{7}$$

and finally for the shear stress:

$$\tau = \frac{T}{2\pi R(2H_0 - D)} \approx \frac{8\pi\eta_s V}{s\left(1 - \frac{D}{2H_0}\right)}\left[\left(\frac{2H_0}{D}\right)^{\frac{3}{4}} - 1\right] \tag{8}$$

We compare on Fig. 7b such prediction with the data obtained for sample S5, for which the range of compression over which low friction is observed is the largest. Using $V = 1\ \mu m.s^{-1}$, $s \approx 16\ nm$ (obtained from the fit of the normal force curve), we obtain the correct order of magnitude for the shear stress by setting $\eta_s \approx 0.022\ Pa.s$. However, we can see on Fig. 7b that the dependence of $\tau$ on the compression ratio $2H_0/D$ is not properly captured by the model, the experimentally observed power law in the low friction regime having an exponent larger than the predicted 3/4 value.

We propose a toy model in line with what was described in a previous SFA study of sheared adsorbed polymer layers [13]. We assume that the response of the confined adsorbed chitosan layers in the "low friction" regime is that of a semi-dilute solution of unentangled chains containing $N_b$ monomers in good solvent. The viscosity of such semi-dilute solution is given by [46]:

$$\eta \approx 6\pi\eta_s N_b \phi^{\frac{5}{4}} \tag{9}$$

with $\phi$ the monomer volume fraction. We assume that the monomer volume fraction of the unperturbed layers is uniform across their thickness and is given by $\phi_0 = N_b a^3/(s^2 H_0)$, with $s$ the effective distance between chains determined previously, and $a^3$ the monomer volume. We can thus relate the monomer volume fraction in the confined layers to the compression ratio as $\phi/\phi_0 = 2H_0/h$, with $h$ the intersurface distance as in Eq. 5. At the scaling level, we write that the number of monomers, the layer thickness and the interchain distance are related, as for an Alexander-de Gennes brush [35], as $N_b \approx H_0 s^{2/3} a^{-5/3}$. Substituting these relationships in Eq. 9, we finally get for the viscous shear stress between two planar surfaces:

$$\sigma_v = \eta \frac{V}{h} \approx 3\pi \eta_s \frac{V}{s} \left(\frac{2H_0}{h}\right)^{\frac{9}{4}} \quad (10)$$

The above expression for the shear stress is similar to that established by Klein [45], but with a stronger power dependence on the compression ratio. Integrating $\sigma_v$ as done before to obtain the shear force in the sphere/flat geometry and dividing by the contact area, we finally get for the average shear stress:

$$\tau \approx \frac{12\pi \eta_s V}{5s \left(1 - \frac{D}{2H_0}\right)} \left[\left(\frac{2H_0}{D}\right)^{\frac{5}{4}} - 1\right] \quad (11)$$

Using Eq. 11 with $V = 1 \, \mu m.s^{-1}$, $s \approx 16 \, nm$ and $\eta_s \approx 0.025 \, Pa.s$, we see on Fig. 7b that our toy model accounts for the correct dependence of the shear stress on compression ratio over the entire low friction regime of sample S5. This strengthens the fact that the low friction regime exhibited by the confined chitosan layers is indeed a liquid-like regime controlled by viscous dissipation.

However, a quantitative description of the shear stress requires the use of an effective viscosity that is about one order of magnitude larger than that of the aqueous solvent used here. At present, we do not have a clear explanation for such value of the effective viscosity. Moreover, the above model, similarly to the one proposed by Klein, does not account for any explicit dependence on chain length of the magnitude of the shear stress. Describing quantitatively the frictional stress for samples other than S5 would therefore require adjusting further the effective viscosity in the model, with larger values of $\eta_s$ for shorter chain lengths. Accounting for such molecular weight dependence of $\eta_s$ is beyond the scope of the present study and calls for more refined modelling of the shear response of confined polymer layers.

Finally, the transition to the high friction solid-like regime is likely to be related to the dehydration of the layers under strong confinement [47]. The exact mechanism by which such frictional transition takes place, and its observed dependence on chitosan molecular weight, remains to be understood at that stage.

# Conclusion

Chitosan from fungal origin is attracting increasing interest as a biosourced polymer in applications requiring polycationic macromolecules. In cosmetics, it appears as an interesting model system allowing to investigate how a biosourced macromolecule may perform compared to those used in current haircare product formulations. Here we have shown that, in spite of the rather crude properties of our chitosan samples, an impact on intersurface forces of the average molecular weight of the adsorbed chains could be clearly evidenced. Moreover, we have demonstrated that models of steric repulsion or friction established for brushes of neutral polymers could be employed to account semi-quantitatively for our results. Interestingly, we have shown that chitosan films under shear exhibit a frictional transition: at low compression levels, films behave viscouslike, while they display higher, and solid-like, friction force when confined further. Such a transition appears to be controlled by the hydration level of the films, and depends on chain molecular weight, with longer chains able to sustain larger compressions before exhibiting high friction.

# Acknowledgements

We are grateful to Elie Raphaël from Gulliver Lab at ESPCI Paris for stimulating discussions regarding modeling of interaction forces. Thanks to Fabien Frogneaux (L'OREAL) for his support on this project. Financial support from L'Oréal is acknowledged via contract 22CUF10180 with UGA/Floralis.

# Conflict of Interest

L.B. declares no competing financial interest. G.S.L and S.T. are employees of L'OREAL engaged in research activities.

# References

[1] G. S. Luengo, F. Leonforte, A. Greaves, R. G. Rubio, and E. Guzman, Physico-chemical challenges on the self-assembly of natural and bio-based ingredients on hair surfaces: towards sustainable haircare formulations, Green Chem. **25**, 7863 (2023).
[2] G. S. Luengo, A.-L. Fameau, F. Léonforte, and A. J. Greaves, Surface science of cosmetic substrates, cleansing actives and formulations, Adv. Colloid Interface Sci. **290**, 102383 (2021).
[3] C. Bouillon and J. Wilkinson, editors , *The Science of Hair Care*, 2nd ed. (CRC Press, 2005).
[4] C. R. Robbins, *The Physical Properties and Cosmetic Behavior of Hair*, in *Chemical and Physical Behavior of Human Hair* (Springer New York, New York, NY, 1994), pp. 299–370.
[5] S. Llamas, E. Guzmán, F. Ortega, N. Baghdadli, C. Cazeneuve, R. G. Rubio, and G. S. Luengo, Adsorption of polyelectrolytes and polyelectrolytes-surfactant mixtures at surfaces: a physico-chemical approach to a cosmetic challenge, Adv. Colloid Interface Sci. **222**, 461 (2015).
[6] S. Tokunaga, H. Tanamachi, and K. Ishikawa, Degradation of Hair Surface: Importance of 18-MEA and Epicuticle, Cosmetics **6**, 31 (2019).
[7] B. J. Coscia, J. C. Shelley, A. R. Browning, J. M. Sanders, R. Chaudret, R. Rozot, F. Léonforte, M. D. Halls, and G. S. Luengo, Shearing friction behaviour of synthetic polymers compared to a functionalized polysaccharide on biomimetic surfaces: models for the prediction of performance of eco-designed formulations, Phys. Chem. Chem. Phys. **25**, 1768 (2023).
[8] E. Weiand, J. P. Ewen, P. H. Koenig, Y. Roiter, S. H. Page, S. Angioletti-Uberti, and D. Dini, Coarse-grained molecular models of the surface of hair, Soft Matter **18**, 1779 (2022).
[9] E. Weiand, F. Rodriguez-Ropero, Y. Roiter, P. H. Koenig, S. Angioletti-Uberti, D. Dini, and J. P. Ewen, Effects of surfactant adsorption on the wettability and friction of biomimetic surfaces, Phys. Chem. Chem. Phys. **25**, 21916 (2023).
[10] H. J. Taunton, C. Toprakcioglu, L. J. Fetters, and J. Klein, Forces between surfaces bearing terminally anchored polymer chains in good solvents, Nature **332**, 712 (1988).
[11] J. Klein, D. Perahia, and S. Warburg, Forces between polymer-bearing surfaces undergoing shear, Nature **352**, 143 (1991).
[12] J. Klein, E. Kumacheva, D. Mahalu, D. Perahia, and L. J. Fetters, Reduction of frictional forces between solid surfaces bearing polymer brushes, Nature **370**, 634 (1994).
[13] U. Raviv, R. Tadmor, and J. Klein, Shear and Frictional Interactions between Adsorbed Polymer Layers in a Good Solvent, J. Phys. Chem. B **105**, 8125 (2001).
[14] P. M. Claesson, E. Poptoshev, E. Blomberg, and A. Dedinaite, Polyelectrolyte-mediated surface interactions, Adv. Colloid Interface Sci. **114–115**, 173 (2005).

[15] Y. Kamiyama and J. Israelachvili, Effect of pH and salt on the adsorption and interactions of an amphoteric polyelectrolyte, Macromolecules **25**, 5081 (1992).

[16] W. Lin and J. Klein, Recent Progress in Cartilage Lubrication, Adv. Mater. **33**, 2005513 (2021).

[17] U. Raviv, S. Giasson, N. Kampf, J.-F. Gohy, R. Jérôme, and J. Klein, Lubrication by charged polymers, Nature **425**, 163 (2003).

[18] N. M. Harvey, G. E. Yakubov, J. R. Stokes, and J. Klein, Normal and Shear Forces between Surfaces Bearing Porcine Gastric Mucin, a High-Molecular-Weight Glycoprotein, Biomacromolecules **12**, 1041 (2011).

[19] M. Chen, W. H. Briscoe, S. P. Armes, and J. Klein, Lubrication at Physiological Pressures by Polyzwitterionic Brushes, Science **323**, 1698 (2009).

[20] B. Zappone, M. Ruths, G. W. Greene, G. D. Jay, and J. N. Israelachvili, Adsorption, Lubrication, and Wear of Lubricin on Model Surfaces: Polymer Brush-Like Behavior of a Glycoprotein, Biophys. J. **92**, 1693 (2007).

[21] J. Yu, N. E. Jackson, X. Xu, Y. Morgenstern, Y. Kaufman, M. Ruths, J. J. De Pablo, and M. Tirrell, Multivalent counterions diminish the lubricity of polyelectrolyte brushes, Science **360**, 1434 (2018).

[22] P. M. Claesson and B. W. Ninham, pH-dependent interactions between adsorbed chitosan layers, Langmuir **8**, 1406 (1992).

[23] N. Kampf, U. Raviv, and J. Klein, Normal and Shear Forces between Adsorbed and Gelled Layers of Chitosan, a Naturally Occurring Cationic Polyelectrolyte, Macromolecules **37**, 1134 (2004).

[24] Y. Kan, Q. Yang, Q. Tan, Z. Wei, and Y. Chen, Diminishing Cohesion of Chitosan Films in Acidic Solution by Multivalent Metal Cations, Langmuir **36**, 4964 (2020).

[25] Q. Tan, Y. Kan, H. Huang, W. Wu, and X. Lu, Probing the Molecular Interactions of Chitosan Films in Acidic Solutions with Different Salt Ions, Coatings **10**, 1052 (2020).

[26] L. Xiang, L. Gong, J. Zhang, L. Zhang, W. Hu, W. Wang, Q. Lu, and H. Zeng, Probing molecular interactions of PEGylated chitosan in aqueous solutions using a surface force apparatus, Phys. Chem. Chem. Phys. **21**, 20571 (2019).

[27] H. Kwon, J. Choi, C. Lim, J. Kim, A. Osman, Y. Jho, D. S. Hwang, and D. W. Lee, Strong Hydrophobic Interaction of High Molecular Weight Chitosan in Aqueous Solution, Biomacromolecules **26**, 1012 (2025).

[28] E. Claverie, M. Perini, R. C. A. Onderwater, S. Pianezze, R. Larcher, S. Roosa, B. Yada, and R. Wattiez, Multiple Technology Approach Based on Stable Isotope Ratio Analysis, Fourier Transform Infrared Spectrometry and Thermogravimetric Analysis to Ensure the Fungal Origin of the Chitosan, Molecules **28**, 4324 (2023).

[29] S. Krake, C. Conzelmann, S. Heuer, M. Dyballa, S. Zibek, and T. Hahn, Production of chitosan from Aspergillus niger and quantitative evaluation of the process using adapted analytical tools, Biotechnol. Bioprocess Eng. **29**, 942 (2024).

[30] L. Bureau, Surface force apparatus for nanorheology under large shear strain, Rev. Sci. Instrum. **78**, 065110 (2007).

[31] M. Heuberger, The extended surface forces apparatus. Part I. Fast spectral correlation interferometry, Rev. Sci. Instrum. **72**, 1700 (2001).

[32] A. V. Dobrynin, A. Deshkovski, and M. Rubinstein, Adsorption of Polyelectrolytes at Oppositely Charged Surfaces, Macromolecules **34**, 3421 (2001).

[33] D. Lupa, W. Plaziński, A. Michna, M. Wasilewska, P. Pomastowski, A. Gołębiowski, B. Buszewski, and Z. Adamczyk, Chitosan characteristics in electrolyte solutions: Combined molecular dynamics modeling and slender body hydrodynamics, Carbohydr. Polym. **292**, 119676 (2022).

[34] D. W. Schaefer, J. F. Joanny, and P. Pincus, Dynamics of Semiflexible Polymers in Solution, Macromolecules **13**, 1280 (1980).

[35] S. Attili, O. V. Borisov, and R. P. Richter, Films of End-Grafted Hyaluronan Are a Prototype of a Brush of a Strongly Charged, Semiflexible Polyelectrolyte with Intrinsic Excluded Volume, Biomacromolecules **13**, 1466 (2012).

[36] P. G. De Gennes, Polymers at an interface. 2. Interaction between two plates carrying adsorbed polymer layers, Macromolecules **15**, 492 (1982).

[37] M. Aubouy, O. Guiselin, and E. Raphaël, Scaling Description of Polymer Interfaces: Flat Layers, Macromolecules **29**, 7261 (1996).

[38] S. Block and C. A. Helm, Single Polyelectrolyte Layers Adsorbed at High Salt Conditions: Polyelectrolyte Brush Domains Coexisting with Flatly Adsorbed Chains, Macromolecules **42**, 6733 (2009).
[39] A. Tiraferri, P. Maroni, D. Caro Rodríguez, and M. Borkovec, Mechanism of Chitosan Adsorption on Silica from Aqueous Solutions, Langmuir **30**, 4980 (2014).
[40] S. Cozzolino, P. Gutfreund, A. Vorobiev, R. J. L. Welbourn, A. Greaves, F. Zuttion, M. W. Rutland, and G. S. Luengo, Adsorption hierarchy of surfactants and polymers to a damaged hair model: effect of composition, order and polymer size, Phys. Chem. Chem. Phys. **27**, 1089 (2025).
[41] U. Raviv, J. Klein, and T. A. Witten, The polymer mat: Arrested rebound of a compressed polymer layer, Eur. Phys. J. E **9**, 405 (2002).
[42] L. Bureau, T. Baumberger, and C. Caroli, Rheological aging and rejuvenation in solid friction contacts, Eur. Phys. J. E **8**, 331 (2002).
[43] K. L. Johnson, *Contact Mechanics*, 1st ed. (Cambridge University Press, 1985).
[44] R. G. Horn, J. N. Israelachvili, and F. Pribac, Measurement of the deformation and adhesion of solids in contact, J. Colloid Interface Sci. **115**, 480 (1987).
[45] J. Klein, Shear, Friction, and Lubrication Forces Between Polymer-Bearing Surfaces, Annu. Rev. Mater. Sci. **26**, 581 (1996).
[46] M. Rubinstein and R. H. Colby, *Polymer Physics* (Oxford University PressOxford, 2003).
[47] J. Klein, E. Kumacheva, D. Perahia, and L. J. Fetters, Shear forces between sliding surfaces coated with polymer brushes: The high friction regime, Acta Polym. **49**, 617 (1998).

---

[i] To circumvent the noise sensitivity of taking the derivative of an experimental signal exhibiting scatter, we proceed as follows to compute the local friction coefficient. In the high (linear) friction regime, µ is taken as the slope of the best affine fit to the data. In the low (non-linear) friction regime, we fit our data with a smooth and monotonous polynomial function and take the derivative with respect to normal load of this fit function as the value of µ for each applied load.

[ii] Such an estimate neglects the electrostatic part of the excluded volume [32], $v_{elec} \approx f^2 \lambda_D^2 l_B$, which in our conditions is ~1.6 nm, compared to the excluded volume of a semi-flexible chain, given by [34] $v \approx l_p^2 a \approx 12.5\ nm$ with $l_p = na = 5\ nm$ for chitosan [33].

[iii] We assume here that the adsorbed layers behave as Winkler's elastic foundations [43], and we neglect the Hertz-like elastic deformation of the mica sheets and underlying glue. Such an assumption is justified by the fact that, computing the contact area of the Hertz problem in the range of loads applied here, and using for the effective elastic modulus of the mica/glue layers a value [44] of 5 GPa, we indeed find that it is much smaller than the contact area computed from layer indentation (between 100 and 4 times smaller depending on load and adsorbed layer stiffness).

# Surface forces and frictional properties of adsorbed chitosan films in salted aqueous medium

*Lionel Bureau[1], Simon Trouillé[2] and Gustavo S. Luengo[2]*

[1]Université Grenoble Alpes, CNRS, LIPhy, 38000, Grenoble, France
[2]L'Oréal Research and Innovation, Aulnay-Sous-Bois, France

# Supplementary information

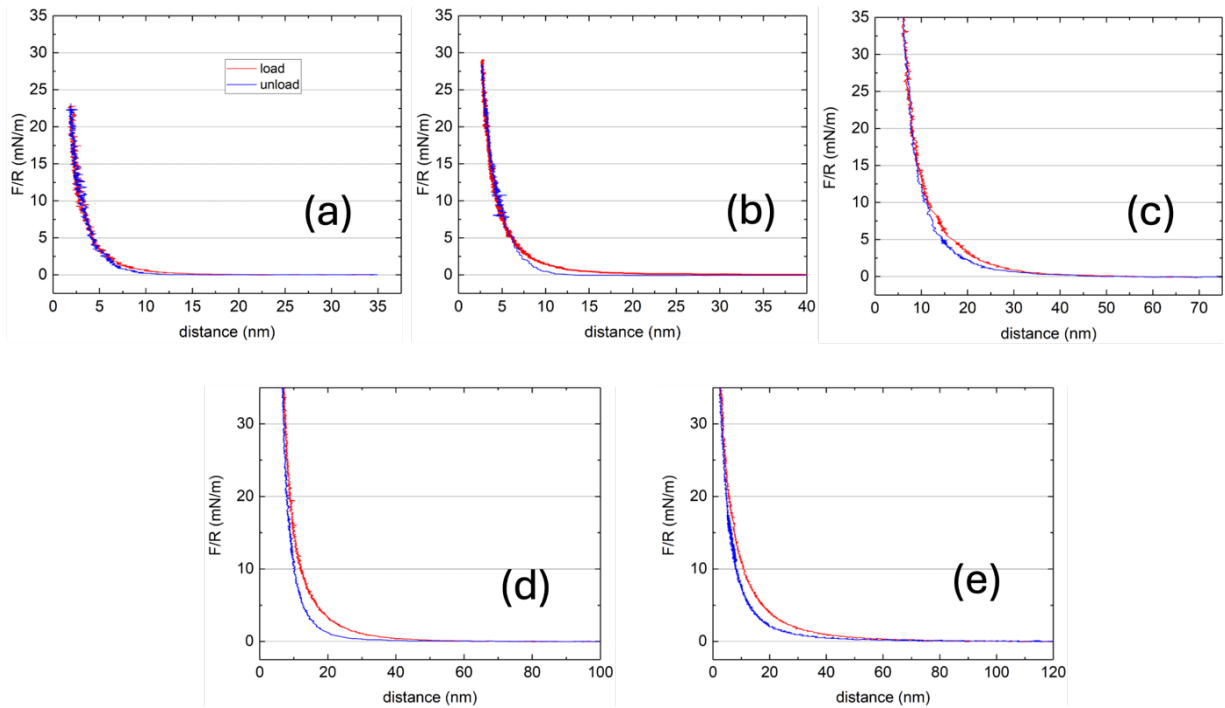

*Fig. S1. Load (red) and unload (blue) curves for sample S1 (a), S2 (b), S3 (c), S4 (d) and S5 (e).*

The Alexander-de Gennes model for brush compression was also used to fit the compression part of the experimental force distance curves. Within this framework, the force F is predicted to vary with distance D as:

$$\frac{F(D)}{R} \approx \frac{2H_0 kT}{s^3}\left[7\left(\frac{2H_0}{D}\right)^{5/4} + 5\left(\frac{D}{2H_0}\right)^{7/4} - 12\right]$$

As described in the main text, we fit each individual force curve with the above equation, for F/R<1 mN/m, to obtain values of $H_0$ and $s$, and plot the rescaled force $F \times (s^3)/(R(2H_0)\,kT)$ as a function of rescaled distance $D/(2H_0)$. As seen on Fig. S2, such a rescaling provides again a satisfactory collapse of all the datasets. Comparison of the experimental data with the functional form of Alexander-de Gennes theoretical expression shows that force curves are well described by polymer brush theory in the range $D/(2H_0) \geq 0.4$, similarly to SCFT. $H_0$ is found to be proportional to $R_F$ for the 5 molecular weights investigated. Parameter $s$ is found to be insensitive to chitosan molecular weight and lies in the range $s=10-16$ nm.

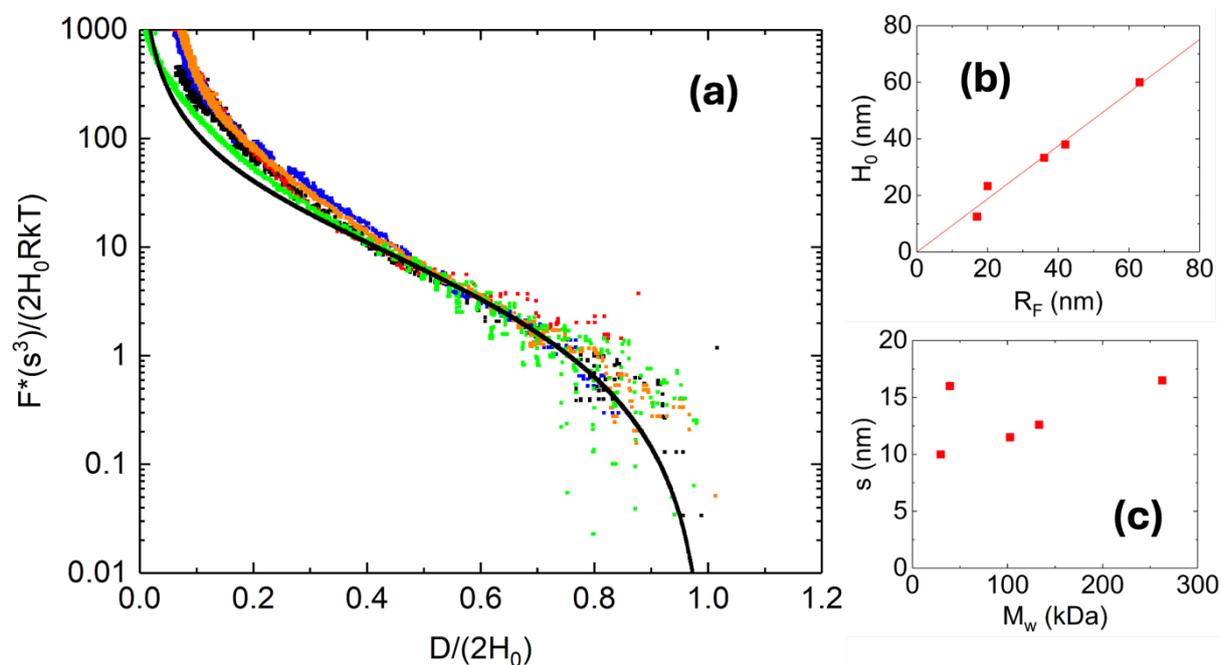

*Fig. S2. (a) Collapsed experimental force curves, with scaled F and D as described in the text (symbols), compared to the theoretical prediction of the Alexander-de Gennes theory of polymer brushes (line). (b) Best fit value for brush thickness $H_0$ as a function of the Flory radius $R_F$ of the chains (symbols). The line is the best linear fit of $H_0(R_F)$, with slope 0.94. (c) Dependence on chain molecular weight $M_w$ of the best fit value of parameter s.*